\documentclass[twocolumn]{elsart3p}

\usepackage{graphicx}
\usepackage{amssymb}
\usepackage{bm}

\begin{document}

\begin{frontmatter}



%% e.g. \dochead{17th International Conference on Dynamical Processes in Excited States of Solids}

\title{Effect of Born and unitary impurity scattering on the Kramer-Pesch shrinkage of a vortex core in an $s$-wave superconductor}


\author[A,C]{Nobuhiko Hayashi}
\author[A,B,C]{Yoichi Higashi}
\author[A,C]{Noriyuki Nakai}
\author[A,C]{Hisataka Suematsu}

\address[A]{NanoSquare Research Center (N2RC), Osaka Prefecture University, 1-2 Gakuen-cho, Naka-ku, Sakai 599-8570, Japan}
\address[B]{Department of Mathematical Sciences, Osaka Prefecture University, 1-1 Gakuen-cho, Naka-ku, Sakai 599-8531, Japan}
\address[C]{CREST(JST), 4-1-8 Honcho, Kawaguchi, Saitama 332-0012, Japan}

%\end{frontmatter}
%\newpage
%\begin{frontmatter}

\begin{abstract}
We theoretically investigate a non-magnetic impurity effect on the temperature dependence of the vortex core shrinkage (Kramer-Pesch effect) in a single-band $s$-wave superconductor.
The Born limit and the unitary limit scattering are compared within the framework of the quasiclassical theory of superconductivity.
We find that the impurity effect inside a vortex core in the unitary limit is weaker than in the Born one
when a system is in the moderately clean regime, which results in a stronger core shrinkage in the unitary limit than in the Born one.
\end{abstract}


\end{frontmatter}

% \linenumbers

\section{Introduction}
%\label{}
The radius of a vortex core in type-II superconductors is one of the fundamental physical quantities which characterize a property of superconductivity.
The temperature and magnetic field dependence of the core radius has been investigated theoretically and experimentally \cite{KP,hayashi05,sonier04,doettinger97,miller00,sonier00,gygi91,ichioka96,volodin97,hayashi98,kato01,gumann06,tanaka07,karmakar10,golubov94,atkinson08,miranovic04,laiho08,belova11}.
%(see Refs.~\cite{hayashi05,sonier04,atkinson08,belova11} and references therein).
The low-temperature vortex core shrinkage, called the Kramer-Pesch (KP) effect \cite{KP}, was theoretically investigated
under the influence of non-magnetic impurities in the Born limit previously \cite{hayashi05}.
%%%
Impurity effects are characterized by the scattering phase shift related to
the impurity potential strength \cite{sigrist91,preosti94,choi89,shift}.
The Born limit corresponds to the limit of weak impurity potential and correspondingly small phase shift.
The opposite limit is called the unitary limit, where
the impurity potential is infinitely strong and the phase shift is $\pi/2$.
The difference between these limits plays an important role in,
for example, unconventional superconductors \cite{sigrist91,preosti94,choi89}.
%%%

In this paper, we theoretically study the KP effect both in the Born and the unitary limit
in an $s$-wave superconductor, and compare their results.
It is found that the temperature dependence of the core shrinkage is stronger in the unitary limit than in the Born one, in the moderately clean regime where the mean free path is of the order of or larger than the coherence length.
%%%
Such a difference of the core shrinkage can be investigated experimentally
by, e.g., muon spin rotation \cite{sonier04,miller00,sonier00}, scanning tunneling microscopy \cite{fitting},
resistivity \cite{doettinger97},
and specific heat \cite{kato01,log}
if there is a suitable superconducting material in which different types and densities of impurities can be doped.

%%%%%%%%%%%%%%%%%%%%%%%%%%%%%%
\section{Formulation}
\begin{figure}[!t]
  \begin{center}
\includegraphics[scale=0.55]{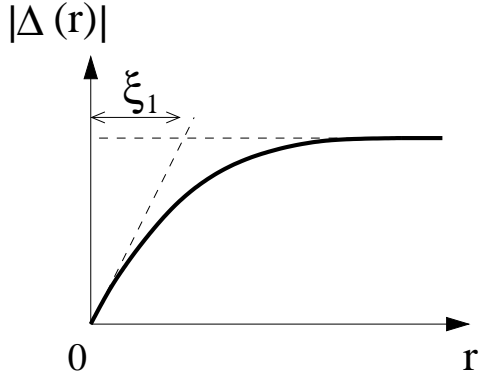}
\caption{  \label{fig:1}
Schematic figure of the pair potential as a function of the distance from the vortex center.
It depicts the length $\xi_1$ that characterizes the vortex core radius [see Eq. (\ref{eq:KP})].
}
  \end{center}
\end{figure}

We consider a single vortex in a single-band $s$-wave superconductor.
The system is assumed to be an isotropic two-dimensional conduction layer perpendicular to the vorticity along the $z$ axis.
In a circular coordinate system within the layer,
the real-space position is ${\bm r}=(r\cos\phi,r\sin\phi)$.
The unit vector ${\bar{\bm k}}$
%=(\cos\theta, \sin\theta)$
represents the sense of the wave number
on a Fermi surface assumed to be circular.
The Fermi velocity is ${\bm v}_{\rm F}=v_{\rm F}{\bar{\bm k}}$.
The pair potential around the vortex is $\Delta({\bm r})=\Delta(r,\phi)=|\Delta(r)| \exp(i\phi)$.
We will consider the temperature $T$ dependence of the length $\xi_1$ that characterizes the vortex core radius \cite{KP,hayashi05,sonier04},
%================================
\begin{equation}
\frac{1}{\xi_1}
=\frac{1}{\Delta(r \rightarrow \infty)}
\lim_{r \rightarrow 0} \frac{\Delta(r)}{r}.
\label{eq:KP}
\end{equation}
%================================
This quantity is depicted in Fig.~\ref{fig:1}.
Note that
$\xi_1$ is related to the pair potential slope at the vortex center
and scales with 
the distance at which the vortex current reaches
its maximum value \cite{KP,sonier04,sonier00,kato01},
while $|\Delta(r)|$ is restored
at a distance of the order of  the coherence length ($\gg\xi_1$ in the clean limit) even at low temperatures \cite{kato01,ovchinnikov}.

%%%%%%%%%
%%%%%%%%%

%%%%%%%%%
\begin{figure}[!t]
  \begin{center}
\includegraphics[scale=0.63]{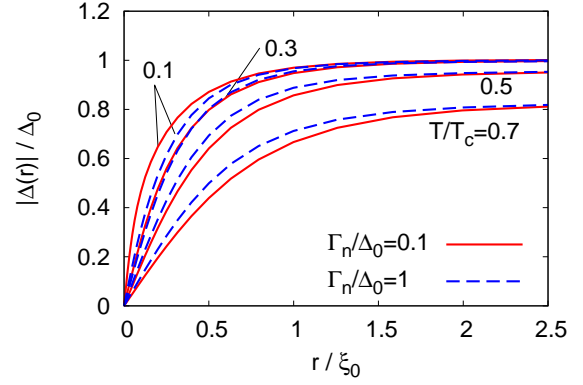}
\caption{  \label{fig:1-2}
Spatial profiles of the pair potential amplitude $|\Delta(r)|$ around the vortex under the influence of impurity scattering in the unitary limit.
The horizontal axis $r$ is the distance from the vortex center.
The scattering rate is $\Gamma_n/\Delta_0=0.1$ (solid lines) and $\Gamma_n/\Delta_0=1$ (dashed lines).
For each scattering rate, the temperature is $T/T_{\mathrm c}=0.1$--0.7
from top to bottom by 0.2 step.
}
  \end{center}
\end{figure}
%%%%%%%%%

%%%%%%%%%
\begin{figure}[!t]
  \begin{center}
\includegraphics[scale=0.63]{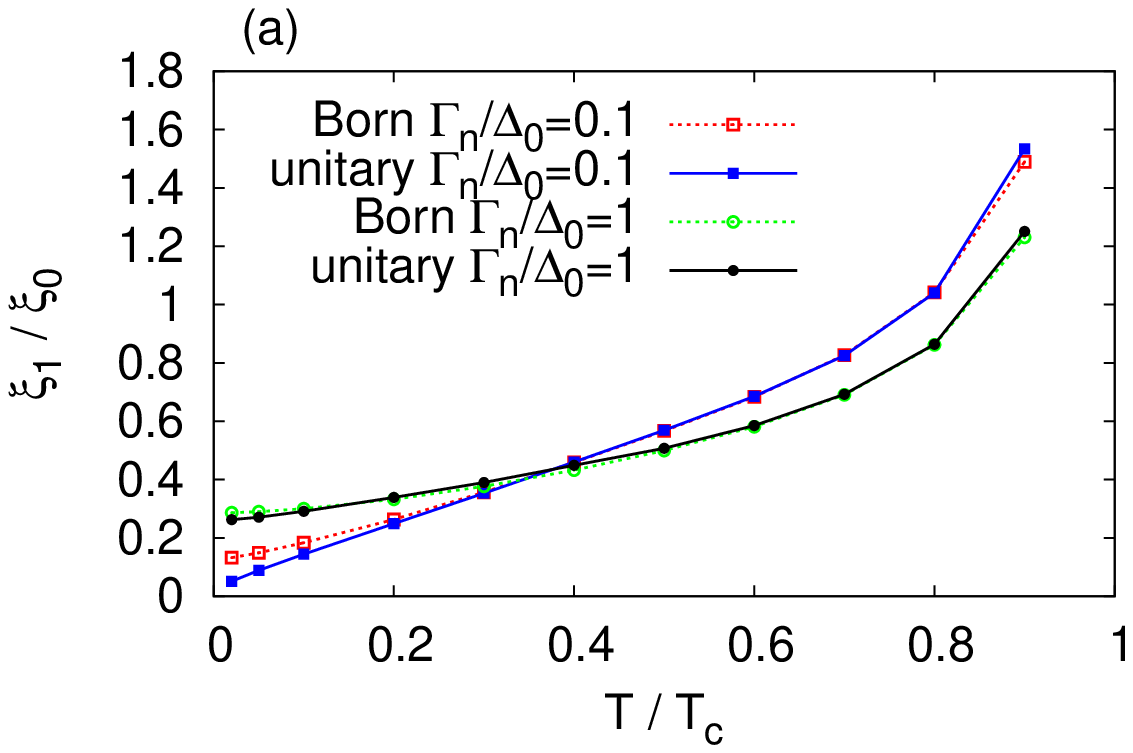}
\includegraphics[scale=0.63]{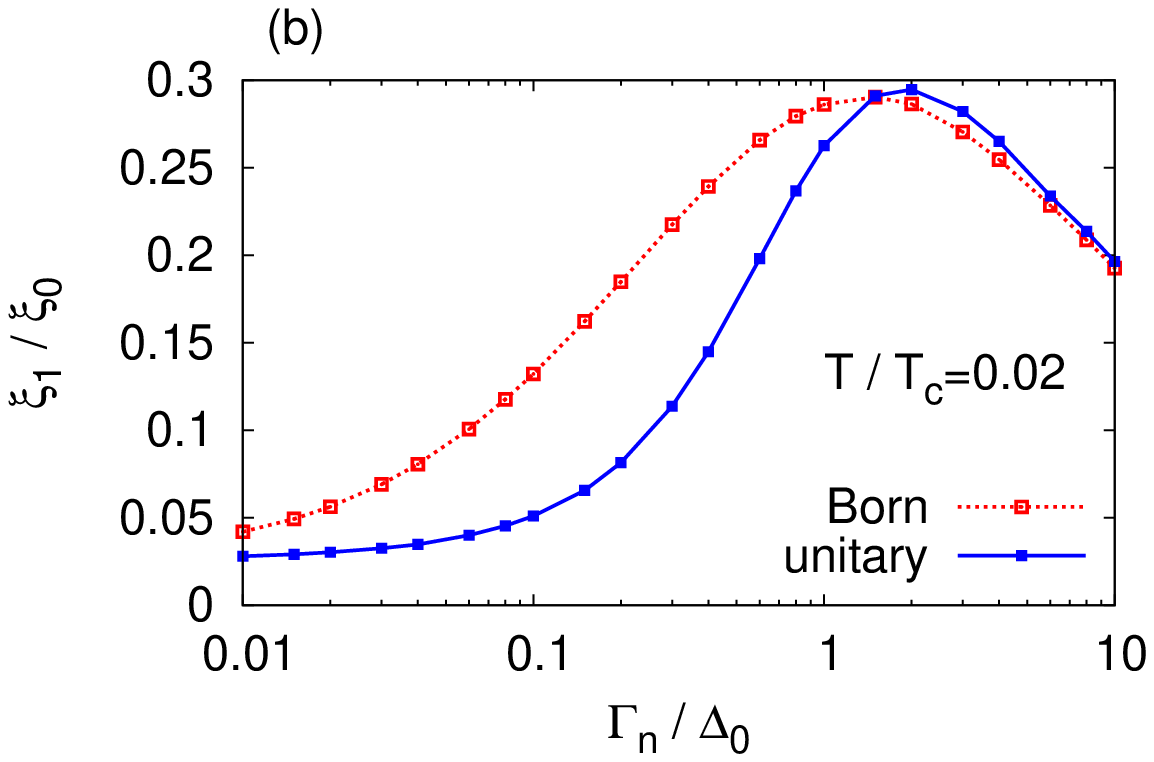}
\caption{  \label{fig:2}
(a) The temperature $T$ dependence of the vortex core radius $\xi_1$.
(b) The scattering rate $\Gamma_n$ dependence of $\xi_1$ at $T/T_{\rm c}=0.02$.
Note that the horizontal axis is in log scale in (b).
The lines are guides for eyes.
}
  \end{center}
\end{figure}
%%%%%%%%%

To obtain $\xi_1(T)$, the vortex core structure is calculated by means of the quasiclassical theory of superconductivity
as in Ref.~\cite{hayashi05}.
%(refer to this reference for details of the following notations).
The Eilenberger equation is numerically solved to obtain the quasiclassical Green's function ${\hat g}(i\omega_n,{\bm r},{\bar{\bm k}})$.
The effect of impurities distributed randomly in the system is taken into account through the impurity self energy ${\hat \Sigma}(i\omega_n,{\bm r},{\bar{\bm k}})$.
The quasiclassical Green's function,
the impurity self energy, and
the Eilenberger equation to be solved are, respectively, given as \cite{hayashi05,sauls09,choi93,eilenberger,rieck93,kusunose}
%================================
\begin{eqnarray}
{\hat g}=
-i\pi
\pmatrix{
g &
if \cr
-if^{\dagger} &
-g \cr
}, \quad
%%%%%
{\hat \Sigma}
&=&
\pmatrix{
\Sigma_{\rm d} &
\Sigma_{12} \cr
\Sigma_{21} &
-\Sigma_{\rm d} \cr
},
%% \qquad
%%%%%
\end{eqnarray}
%================================
%%%%%%
%================================
\begin{eqnarray}
%\\
i {\bm v}_{\rm F} \cdot
{\bm \nabla}{\hat g}
+ \bigl[ i{\tilde \omega}_n {\hat \tau}_{3}-\hat{\tilde \Delta},
{\hat g} \bigr]
&=& 0.
\label{eq:eilen2}
\end{eqnarray}
%================================
The equation is supplemented by the normalization condition
${\hat g}^2=-\pi^2 {\hat \tau}_0$ \cite{eilenberger,schopohl80}.
Here, ${\hat \tau}_3$ is the $z$ component of  the Pauli matrix,
${\hat \tau}_0$ is the unit matrix,
and
the brackets denote the commutator $[{\hat A},{\hat B}]={\hat A}{\hat B}-{\hat B}{\hat A}$.
The Eilenberger equation contains  the renormalized Matsubara frequency (pair potential) ${\tilde \omega}_n$ ($\hat{\tilde \Delta}$) defined by
%================================
\begin{eqnarray}
i{\tilde \omega}_n &=& i\omega_n - \Sigma_{\rm d},
%\quad
\end{eqnarray}
%================================
%%%%%
%================================
\begin{eqnarray}
%\\
\hat{\tilde \Delta} &=&
\pmatrix{
0 &
{\tilde \Delta} \cr
- {\tilde \Delta}^{*}  &
0 \cr
}
=
\pmatrix{
0 &
\Delta + \Sigma_{12} \cr
- (\Delta^{*} - \Sigma_{21}) &
0 \cr
}.
\nonumber
\\
\end{eqnarray}
%================================
We consider an isolated single vortex in an extreme type-II superconductor
(Ginzburg-Landau parameter $\kappa \gg 1$), and therefore the vector potential is
neglected in Eq.~(\ref{eq:eilen2}).
Throughout the paper, we use units in which $\hbar = k_{\rm B} = 1$.

The Eilenberger  equation (\ref{eq:eilen2}) can be solved by the Riccati parametrization \cite{schopohl98,schopohl95,eschrig00}.
The quasiclassical Green's function is expressed as
%%%%%%%%
\begin{eqnarray}
{\hat g}
=
-i\pi
\frac{\operatorname{sgn}(\omega_n)}{1+ab}
\pmatrix{
1-ab &
i2a \cr
-i2b &
-(1-ab) \cr
}.
\end{eqnarray}
%%%%%%%%
Here,
$\operatorname{sgn}(\omega_n)$
is the signum (or sign) function.
The two quantities
$a(i\omega_n,{\bm r},{\bar{\bm k}})$ and $b(i\omega_n,{\bm r},{\bar{\bm k}})$
are independently determined by solving
the Riccati equations,
%%%%%%%%
\begin{eqnarray}
\label{eq:a}
{\bm v}_{\rm F}\cdot{\bm \nabla} a +(2{\tilde \omega}_n +{\tilde \Delta}^* a) a -{\tilde \Delta}    &=& 0, \\
\label{eq:b}
{\bm v}_{\rm F}\cdot{\bm \nabla} b -(2{\tilde \omega}_n +{\tilde \Delta}    b) b +{\tilde \Delta}^* &=& 0.
\end{eqnarray}
%%%%%%%%
These differential equations are solved
along a straight line parallel to ${\bm v}_{\rm F}$
\cite{schopohl98,hayashi97,nagai06}
by using the bulk solutions as initial values \cite{schopohl98,initial},
%%%%%%%%
\begin{eqnarray}
%\label{35}
a_{-\infty}
&=&
\frac{
-{\tilde \omega}_n + \sqrt{{\tilde \omega}_n^2+|{\tilde \Delta}|^2}
}{
{\tilde \Delta}^*
} \quad (\omega_n>0), \\
%%%
%\label{36}
b_{+\infty}
&=&
\frac{
-{\tilde \omega}_n + \sqrt{{\tilde \omega}_n^2+|{\tilde \Delta}|^2}
}{
{\tilde \Delta}
} \quad (\omega_n>0),
\end{eqnarray}
%%%%%%%%
and
%%%%%%%%
\begin{eqnarray}
%\label{37}
a_{+\infty}
&=&
\frac{-1}{b_{+\infty}}
=
\frac{
-{\tilde \omega}_n - \sqrt{{\tilde \omega}_n^2+|{\tilde \Delta}|^2}
}{
{\tilde \Delta}^*
} \quad (\omega_n<0), \\
%%%
%\label{38}
b_{-\infty}
&=&
\frac{-1}{a_{-\infty}}
=
\frac{
-{\tilde \omega}_n - \sqrt{{\tilde \omega}_n^2+|{\tilde \Delta}|^2}
}{
{\tilde \Delta}
} \quad (\omega_n<0).
\end{eqnarray}
%%%%%%%%
A stable numerical solution for $a$ ($b$) is obtained by solving the Riccati equation
in forward (backward) direction along the straight line for $\omega_n>0$ \cite{schopohl98,nagai}.
By contrast, the equation for $a$ ($b$) is solved in backward (forward) direction
for $\omega_n<0$.

Considering an $s$-wave non-magnetic impurity scattering and the $t$-matrix,
${\hat \Sigma}$ is given by \cite{hayashi05,sauls09,preosti96}
%================================
\begin{eqnarray}
{\hat \Sigma}(i\omega_n, {\bm r})
=
\frac{\Gamma_{\rm n} }{1-(\sin^2\delta_0)(1-C)}
\pmatrix{
-i \langle g \rangle &
   \langle f \rangle \cr
 - \langle f^{\dagger} \rangle &
 i \langle g \rangle \cr
},
\nonumber
\\
\label{eq:imp-self2}
\end{eqnarray}
%================================
where $C=\langle g \rangle^2 +  \langle f \rangle \langle f^\dagger \rangle$
with $\langle \cdots \rangle$ being
the average over the Fermi surface with respect to ${\bar{\bm k}}$.
The impurity scattering rate in the normal state is
$\Gamma_{\rm n}$, which is related to the mean free path
$l=v_{\rm F}/2\Gamma_{\rm n}$.
The scattering phase shift is $\delta_0$.
We set $\delta_0=0$ in the Born limit (keeping $\Gamma_{\rm n}$ finite) and $\delta_0=\pi/2$ in the unitary limit.

   The self-consistency equation for $\Delta$,
called the gap equation, is given as
%================================
\begin{equation}
\Delta({\bm r})
=\lambda \pi T
\sum_{-\omega_{\rm c} < \omega_n < \omega_{\rm c}}
\Bigl\langle f(i\omega_n, {\bm r},{\bar {\bm k}})
\Bigr\rangle,
\label{eq:gap}
\end{equation}
%================================
where
$\omega_{\rm c}$ is the cutoff energy
and the coupling constant $\lambda$ is given by
%================================
\begin{equation}
\frac{1}{\lambda}
=
\ln\Bigl(\frac{T}{T_{{\rm c}}} \Bigr)
+ \sum_{0 \le n < (\omega_{\rm c}/\pi T -1)/2}   \frac{2}{2n+1}.
\label{eq:coupling}
\end{equation}
%================================
 Here, $T_{{\rm c}}$ is the superconducting critical temperature.
We set $\omega_{\rm c}=10\Delta_0$
with $\Delta_0$ being the BCS pair-potential amplitude at zero temperature.

The Eilenberger (Riccati) equation, the impurity self energy, and the gap equation 
are numerically solved self-consistently.
The used boundary conditions for the pair potential and impurity self energy
far from the vortex
are the same as those discussed in Ref.~\cite{hayashi05}.
See the Appendix for more details on the calculation procedure.
In the next section, we will show results obtained from self-consistent solutions.
We define the zero-temperature coherence length $\xi_0=v_{\rm F}/\Delta_0$.

%%%%%%%%%%%%%%%%%%%%%%%%%%%%%%

%%%%%%%%%
\begin{figure}[!t]
  \begin{center}
\includegraphics[scale=0.63]{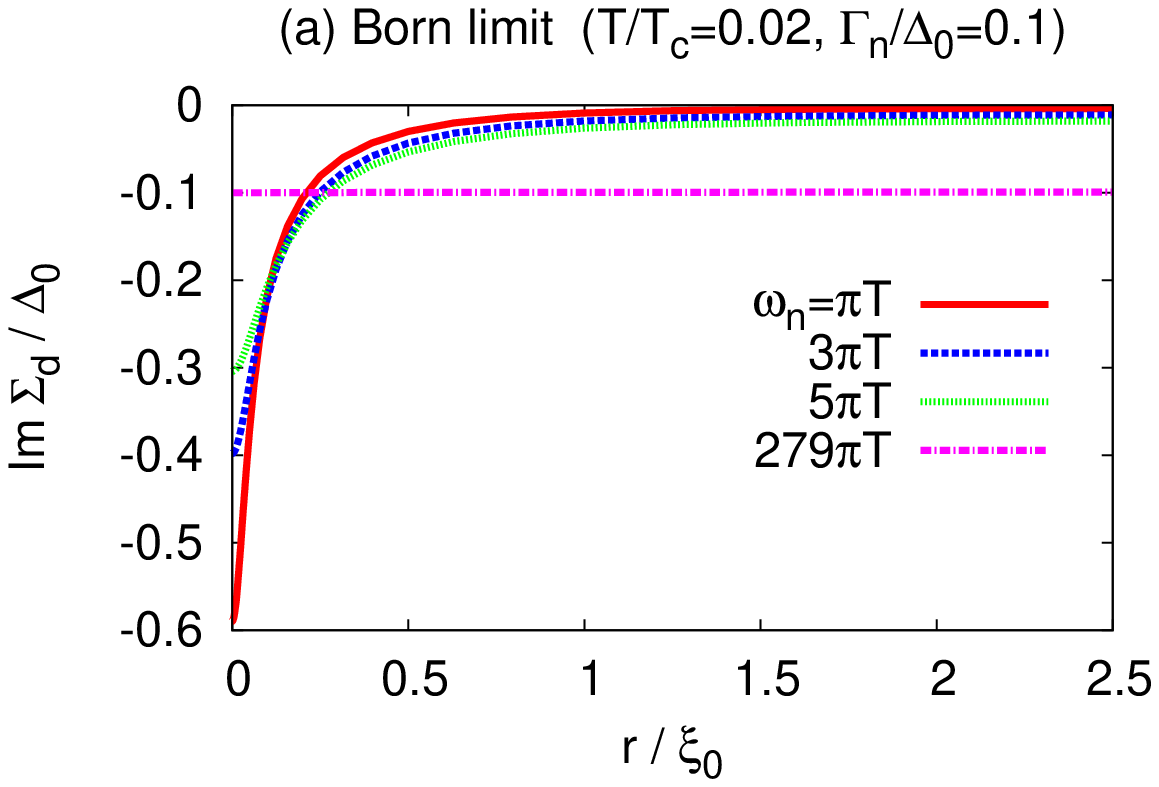}
\includegraphics[scale=0.63]{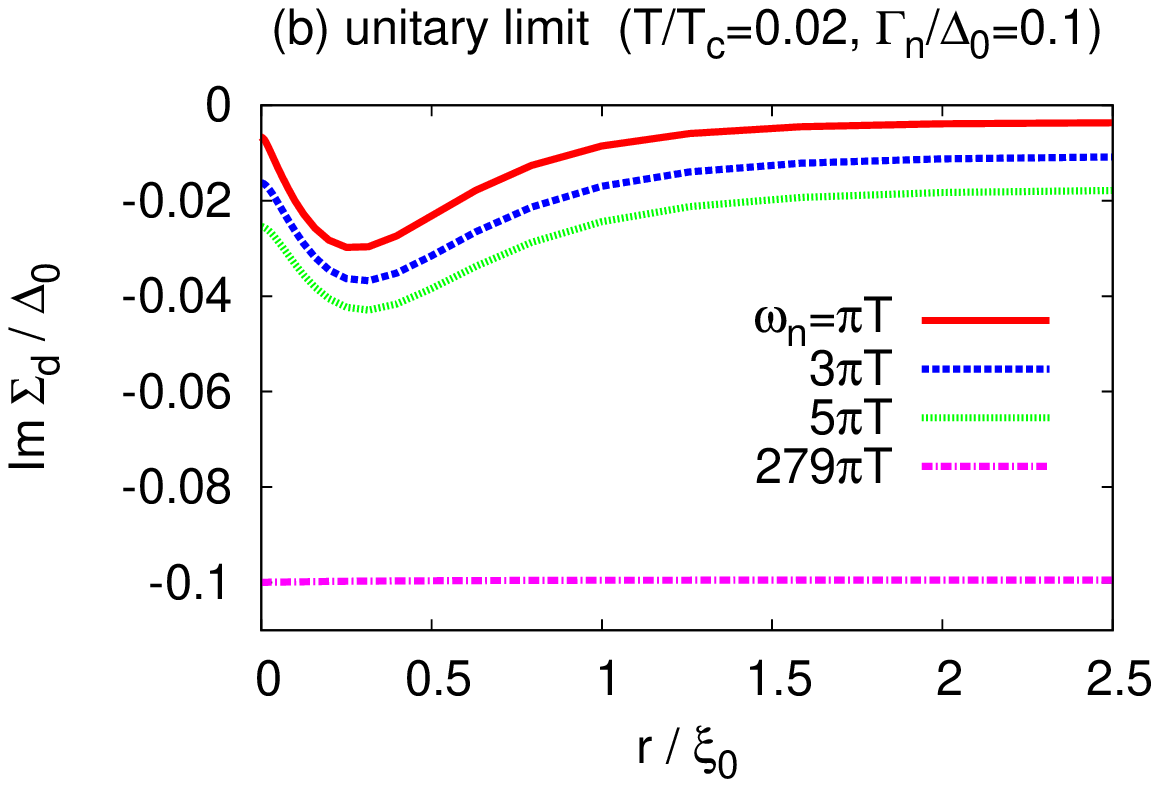}
\caption{  \label{fig:3}
Spatial profiles of the imaginary part of the diagonal impurity self energy $\Sigma_{\rm d}$
for several Matsubara frequencies $\omega_n$ in the Born limit (a) and in the unitary limit (b).
The horizontal axis $r$ is the distance from the vortex center.
Note that $\xi_1/\xi_0 \approx 0.13$ in the Born limit and $\approx 0.05$ in the unitary one
at $T/T_{\rm c}=0.02$ and $\Gamma_n/\Delta_0=0.1$.
}
  \end{center}
\end{figure}
%%%%%%%%%

%%%%%%%%%
\begin{figure}[!t]
  \begin{center}
\includegraphics[scale=0.63]{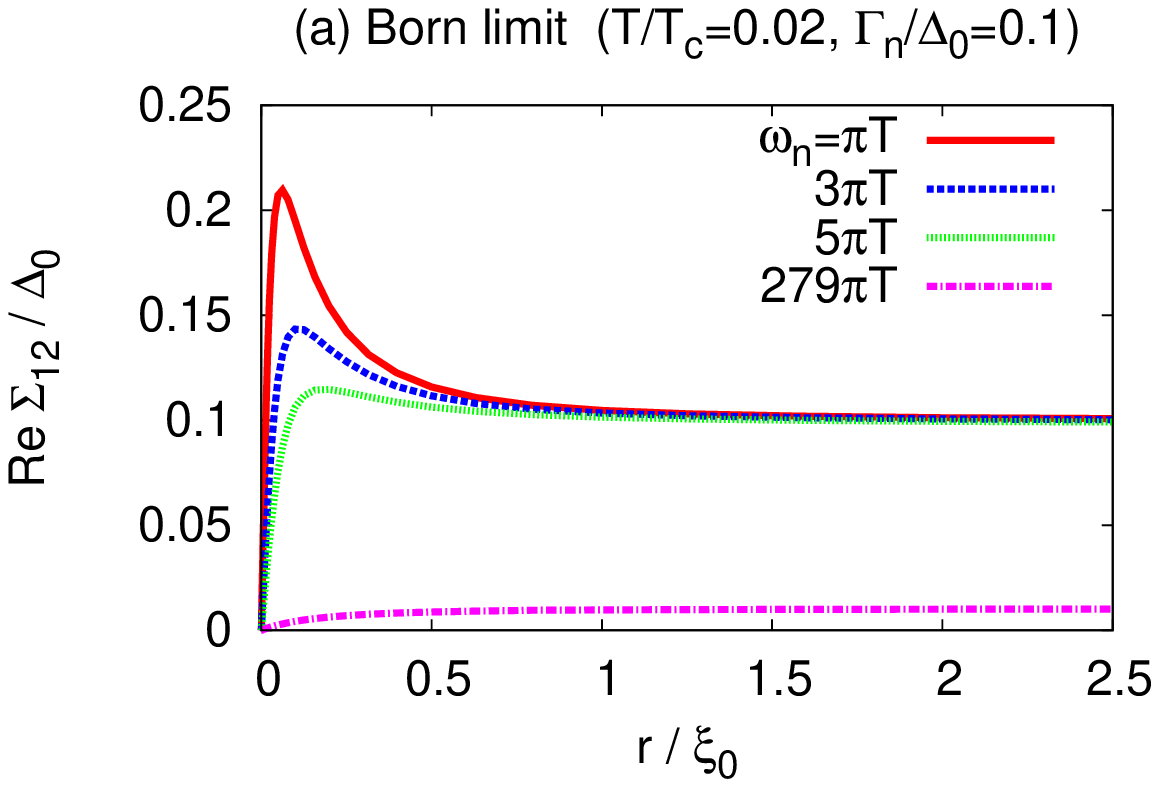}
\includegraphics[scale=0.63]{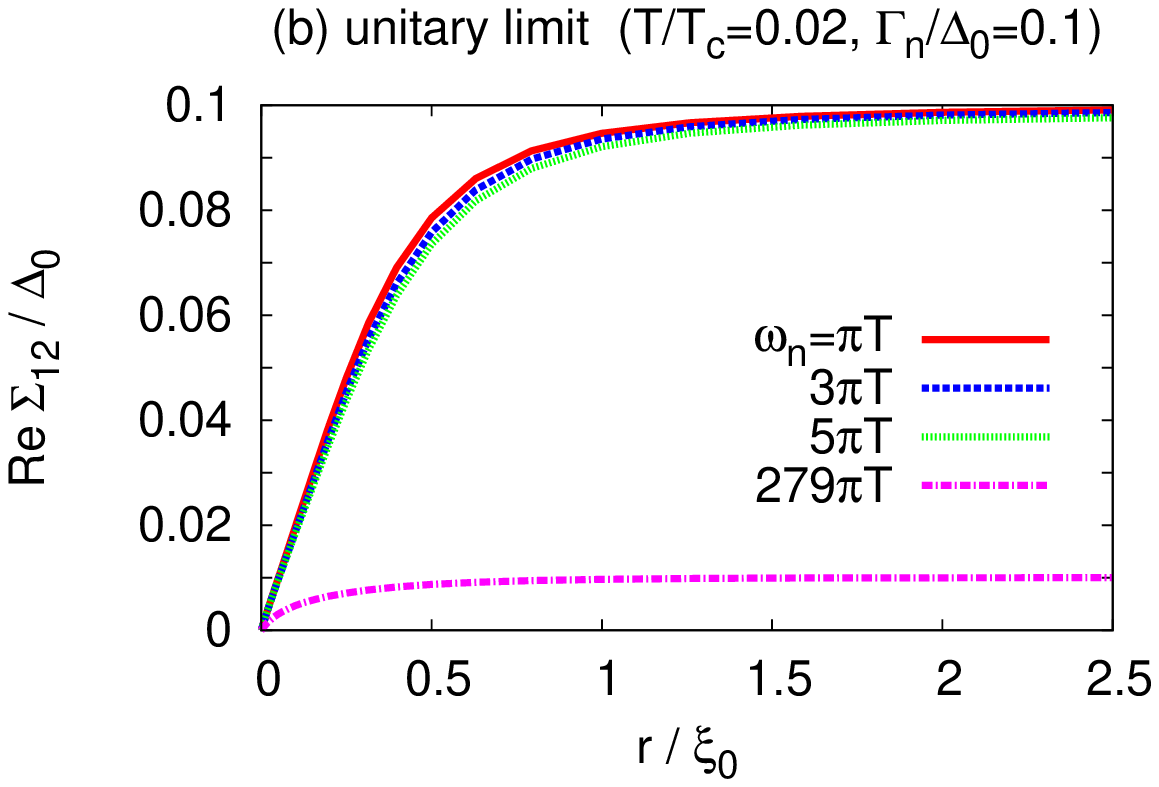}
\caption{  \label{fig:4}
Spatial profiles of the real part of the off-diagonal impurity self energy $\Sigma_{12}$
for several $\omega_n$ in the Born limit (a) and in the unitary limit (b).
$r$ is the distance from the vortex center
along the radial line in the direction $\phi=0$.
}
  \end{center}
\end{figure}
%%%%%%%%%

\section{Result and Discussion}
We show numerical results for the pair potential $\Delta(r)$ in Fig.~\ref{fig:1-2},
which are obtained
under the influence of impurity scattering in the unitary limit.
Similar results are obtained in the Born limit (not shown).
The vortex core radius $\xi_1$ is related to the slope of $\Delta(r)$ at the vortex center $r=0$
according to Eq.~(\ref{eq:KP}) (see also Fig.~\ref{fig:1}).
As seen in Fig.~\ref{fig:1-2},
the slope increases with decreasing the temperature $T$.
The slope becomes smaller with increasing the scattering rate $\Gamma_n$ at the low temperature $T/T_{\mathrm c}=0.1$,
which indicates the suppression of the vortex core shrinkage due to the impurity scattering.
We investigate the impurity effect on the vortex core radius below.

The temperature dependence of the vortex core radius $\xi_1$ [Eq.~(\ref{eq:KP})] is shown in Fig.~\ref{fig:2}(a).
$\xi_1(T)$ is plotted for the scattering rate $\Gamma_n/\Delta_0=0.1$ and $1$ in the Born and unitary limits.
The KP effect means $\xi_1 \propto T$ and $\xi_1 \to 0$ in $T \to 0$ \cite{KP}.
The suppression of  the KP effect, namely the weakened vortex core shrinkage due to impurities, is seen in Fig.~\ref{fig:2}(a).
That is, 
with increasing the scattering rate $\Gamma_n$,
the slope of $\xi_1(T)$ becomes small and
the intercept value of $\xi_1$ in the limit $T \to 0$ becomes large.
Exceptionally, the plot for $\Gamma_n/\Delta_0=0.1$ in the unitary limit seems to exhibit a rather strong KP effect.
To see it in more detail, in Fig.~\ref{fig:2}(b) we plot the dependence of $\xi_1$ on $\Gamma_n$
in the Born and unitary limits at $T/T_{\rm c}=0.02$, which is the lowest temperature in Fig.~\ref{fig:2}(a).
The difference between the Born and unitary limits is clearly seen
in the moderately clean regime ($l \geq \xi_0 \gg l \Delta_0/\varepsilon_{\rm F}$,
namely $ \Delta_0/2\varepsilon_{\rm F} \ll \Gamma_n/\Delta_0 \leq 0.5$; $\varepsilon_{\rm F}$ is the Fermi energy).
This result suggests that inside a vortex core the impurity effect is less effective in the unitary limit than in the Born one
when the system is neither dirty nor too much clean.

As a clue to the difference between the Born and unitary limits,
in Figs.~\ref{fig:3} and \ref{fig:4}
we show the impurity self energies $\Sigma_{\rm d}$ and $\Sigma_{12}$
as functions of the distance $r$ from the vortex center.
Here, we plot $\mathrm{Im}\Sigma_{\rm d}$ and $\mathrm{Re}\Sigma_{12}$ \cite{self} for, as representatives,
several energies $\omega_n=\pi T (2n+1)$ ($n=0, 1, 2$)
and $\omega_{139}=279 \pi T \simeq \omega_{\rm c}$
at $T/T_{\rm c}=0.02$.
The scattering rate is set $\Gamma_n/\Delta_0=0.1$ in common.
We notice that the impurity self energies are enhanced
in the vicinity of the vortex center for small $|\omega_n|$
in the Born limit, but not in the unitary one.

\begin{figure}[!t]
  \begin{center}
\includegraphics[scale=0.63]{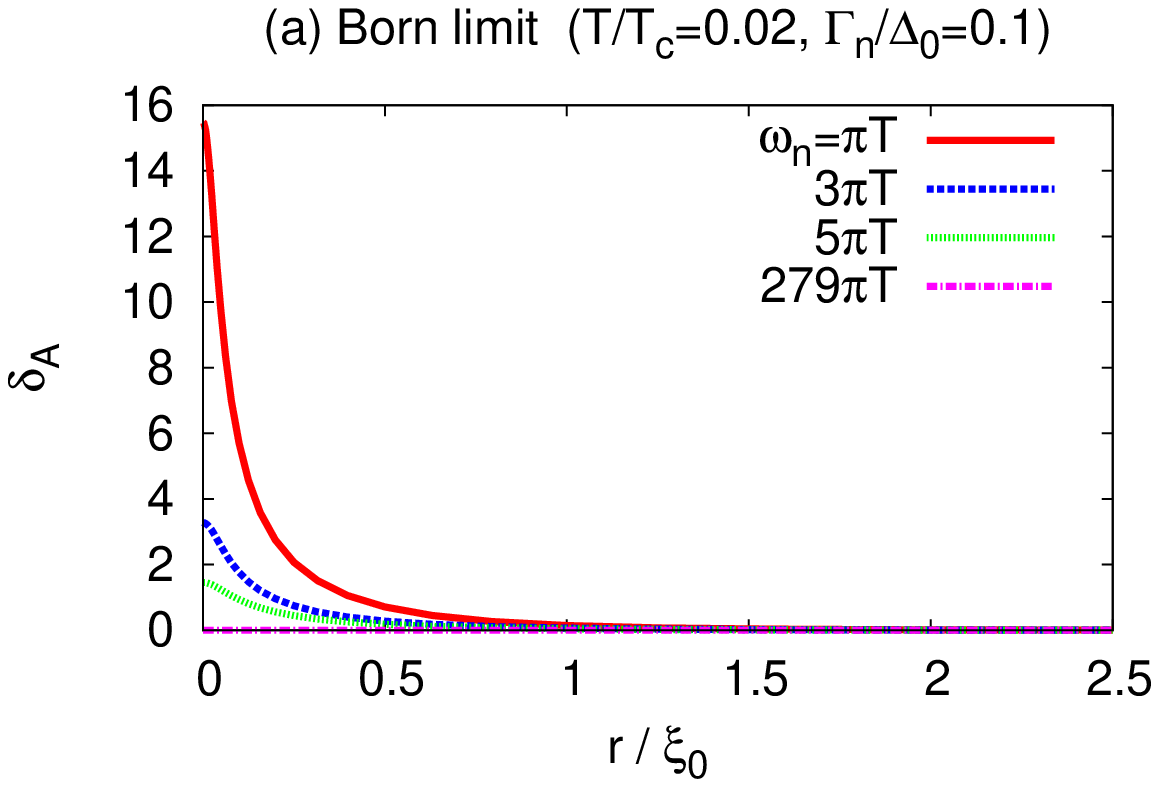}
\includegraphics[scale=0.63]{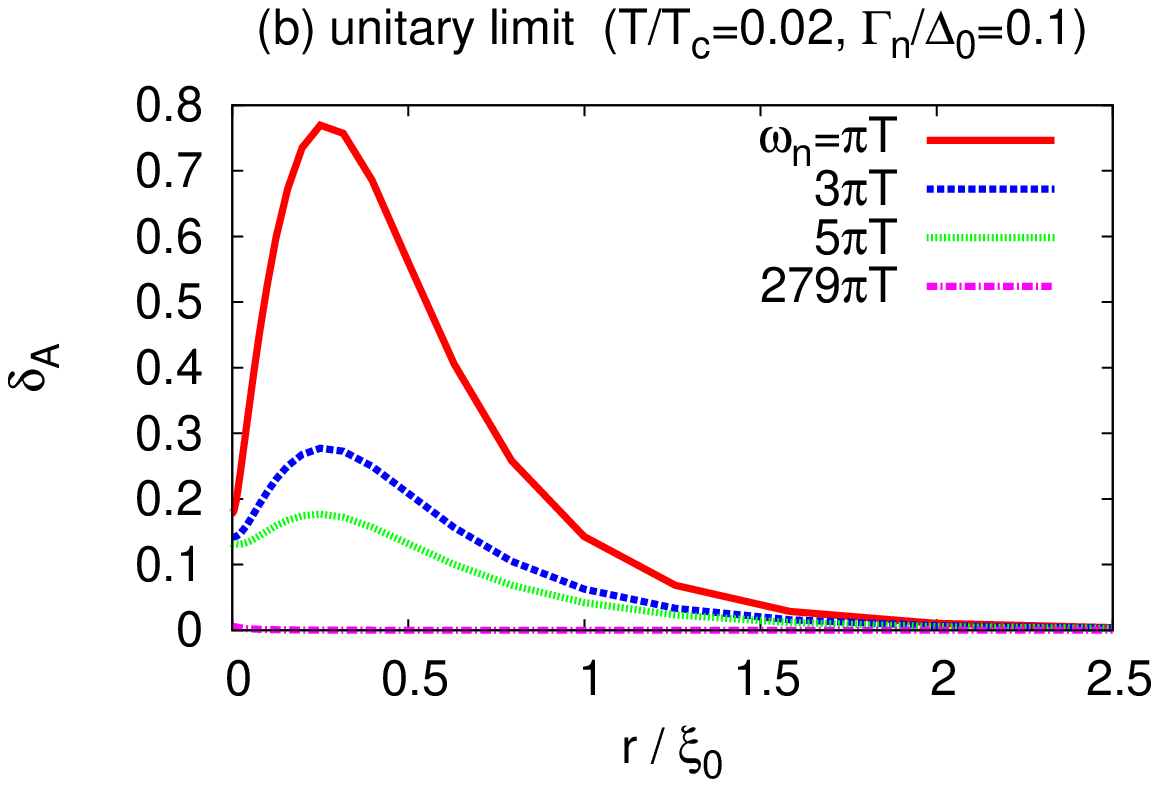}
\caption{  \label{fig:5}
Plots of the indicator $\delta_{\rm A}$ [Eq.~(\ref{eq:deltaA})] as a function of the distance $r$ from the vortex center
for several $\omega_n$  in the Born limit (a) and in the unitary limit (b).
The scattering rate is set $\Gamma_n/\Delta_0=0.1$ in common so as to compare the data between (a) and (b).
}
  \end{center}
\end{figure}

Then, let us investigate the relation between the behavior of the self energies and the impurity effect. 
Impurity effects are ineffective
under the condition that the impurity self energies $\Sigma_{\rm d}$ and $-\Sigma_{12}$ are proportional to $i \omega_n$ and $\Delta$, respectively,
with a common proportionality constant \cite{sigrist91,sign}.
It is the so-called Anderson's theorem \cite{anderson}.
Here, we introduce a dimensionless quantity that measures the degree of the deviation from the above condition:
%================================
\begin{eqnarray}
\delta_{\rm A}
=
\Biggl| \frac{ \Sigma_{\rm d} (i\omega_n,r) }{i\omega_n}
-
         \frac{ -\Sigma_{12} (i\omega_n,r)}{\Delta(r)} \Biggr|.
\label{eq:deltaA}
\end{eqnarray}
%================================
We show $\delta_{\rm A}$ in Fig.~\ref{fig:5},
where $\Gamma_n$ are set the same between the Born and unitary limits.
Indeed, $\delta_{\rm A}$ is zero far away from the vortex center, where Anderson's theorem is satisfied.
In contrast, $\delta_{\rm A}$ becomes finite inside the vortex core.
It is clearly seen that $\delta_{\rm A}$ is prominently large near the vortex center in the Born limit, compared with the unitary-limit case.
This result
indicates that the impurity effect inside the vortex core is weaker in the unitary limit than in the Born one,
resulting in the weaker suppression of the KP effect in the unitary limit (Fig.~\ref{fig:2}).

%%%%%%%%%
%%%%%%%%%

%%%%%%%%%%%%%%%%%%%%%%%%%%%%%%
\section{Conclusion}
We studied the non-magnetic impurity effect on the low-$T$ vortex core shrinkage in the $s$-wave superconductor.
We found that in the moderately clean regime,
the suppression of the core shrinkage is weaker in the unitary limit than in the Born one.
It is attributed to the difference in the impurity self energy between the Born and unitary limits.
However, it is difficult to intuitively understand the reason why the difference appears.
Instead, we introduced the indicator $\delta_{\rm A}$ to estimate the efficiency of the impurity effect.
From the analysis of $\delta_{\rm A}$, it was elucidated that the impurity effect inside the $s$-wave vortex core is stronger in the Born limit than in the unitary one in the moderately clean regime at low temperatures \cite{ichioka}.

%%%%%%%%%%%%%%%%%%%%%%%%%%%%%%%%%%%%%%%%%%%%%%%%%%%%%%%%%%
\appendix
\section*{Appendix}
Here, we briefly explain the calculation procedure on the single vortex.
Our starting point is the Eilenberger equation (\ref{eq:eilen2})
with the impurity self energy and
the impurity-averaged Green's function
(refer to Refs.~\cite{eilenberger,rieck93,kusunose,schopohl80} for its derivation).
It has been used to investigate spatially inhomogeneous systems
such as vortices \cite{hayashi05,miranovic04,laiho08,belova11,sauls09,choi93,kusunose,kato02}.
The Eilenberger  equation is transformed to
the Riccati equations (\ref{eq:a}) and (\ref{eq:b})
(see Refs.~\cite{schopohl98,eschrig00} for the details).
The Eilenberger equation corresponds to three coupled differential equations \cite{thuneberg84},
and therefore it is not so easy to solve.
It is easier to solve the Riccati equations, which are two decoupled differential equations.
Stable solutions are easily obtained when solving the Riccati equations numerically \cite{nagai}.

The Eilenberger and the Riccati equation contain
the differential operator in the form ${\bm v}_{\rm F} \cdot {\bm \nabla}$,
which is treated as follows.
Consider the orthogonal coordinate system
${\bm r}=x \bar{\bm x}$ + $y \bar{\bm y}$ +$z \bar{\bm z}$
with $\bar{\bm z}$ parallel to a rectilinear vortex line.
($\bar{\bm x}$, $\bar{\bm y}$, $\bar{\bm z}$ are orthogonal unit vectors.)
Here, the vortex center is situated on the $z$ axis.
We express the Fermi velocity as
${\bm v}_{\rm F}=v_{\rm F}{\bar{\bm k}}
=v_{\rm F}(\cos\phi_k\sin\theta_k\bar{\bm x}
+\sin\phi_k\sin\theta_k\bar{\bm y}
+\cos\theta_k\bar{\bm z})$.
Because of the translational symmetry along the vortex line,
one can omit the $z$ dependence.
Thus,
${\bm v}_{\rm F} \cdot {\bm \nabla}
\to {\bm v}_{{\rm F}\perp} \cdot {\bm \nabla}
=\bar{\bm x} v_{\rm F}\cos\phi_k\sin\theta_k (\partial /\partial x)
+\bar{\bm y} v_{\rm F}\sin\phi_k\sin\theta_k (\partial /\partial y)$,
where ${\bm v}_{{\rm F}\perp}$ is the Fermi velocity projected onto
the $xy$ plane normal to the vortex line.
When the Fermi surface is a cylinder with central axis parallel to the $z$ axis ($\parallel$ the vortex line),
the angle $\theta_k$ is set to $\pi/2$.
It is convenient to introduce
a new orthogonal coordinate system where
one of the axes is parallel to ${\bm v}_{{\rm F}\perp}$.
Using such coordinates $(s,t)$,
${\bm r}=x \bar{\bm x}$ + $y \bar{\bm y}
=s \bar{\bm s}$ + $t \bar{\bm t}$
with $\bar{\bm s} \parallel {\bm v}_{{\rm F}\perp}$.
The variable $t$ is called the impact parameter.
Eventually, 
${\bm v}_{\rm F} \cdot {\bm \nabla}
=v_{\rm F} \sin\theta_k (\partial /\partial s)$
in the equations to be solved.
One can obtain the quasiclassical Green's function
at a position ${\bm r}_0=x_0 \bar{\bm x}$ + $y_0 \bar{\bm y}=s_0 \bar{\bm s}$ + $t_0 \bar{\bm t}$
by solving the first-order differential equations along a straight line parallel to the $s$ axis with fixing $t=t_0$.
The ${\bar{\bm k}}$ dependence is obtained by solving the equations in the same way
for each different direction $\bar{\bm s}$ and polar angle $\theta_k$.
Note that $\bar{\bm s}$ is a function of the azimuth angle $\phi_k$ of ${\bar{\bm k}}$.

A single vortex is characterized by the pair potential
$\Delta({\bm r})=|\Delta(r)| \exp(i\phi)$
in the case of an $s$-wave superconductor.
Because of the rotational symmetry around the vortex line,
the dependence on
the real-space azimuth angle $\phi$ is factored out as 
$f={\bar f}\exp(i\phi)$,
$f^\dagger={\bar f}^\dagger\exp(-i\phi)$,
$g={\bar g}$,
$\Sigma_{12}={\bar \Sigma}_{12}\exp(i\phi)$,
$\Sigma_{21}={\bar \Sigma}_{21}\exp(-i\phi)$,
$\Sigma_{\rm d}={\bar \Sigma}_{\rm d}$,
$a={\bar a}\exp(i\phi)$,
and
$b={\bar b}\exp(-i\phi)$,
where the quantities with ``bar" are independent of $\phi$.
Therefore, the task is to determine the radial $r$ dependence.

In numerical calculations, the pair potential and the impurity self energies
are represented by their values
at discrete points on the radial line.
The values between those positions
are calculated by linear interpolation.
It is necessary to take dense point spacing
near the vortex center because the values have rapid spatial variations there,
while sparse points are sufficient far from a vortex.
In consideration of it,
we take non-equally-spaced discrete points $r_i$
represented by
$r_i={\tilde r}\bigl[\exp(R_i)-1\bigr]$
with equally-spaced discrete points $R_i$ in dimension-less space.
In the present study,
we set ${\tilde r}=10^{-3}\xi_0$ and take 41 points from the vortex center
to the cutoff distance $r_{\rm c}=10\xi_0$.
In the outside region $r>r_{\rm c}$,
we set $|\Delta(r>r_{\rm c})|=|\Delta(r=r_{\rm c})|$,
${\bar \Sigma}_{12}(r>r_{\rm c})={\bar \Sigma}_{12}(r=r_{\rm c})$,
${\bar \Sigma}_{21}(r>r_{\rm c})={\bar \Sigma}_{21}(r=r_{\rm c})$,
and
${\bar \Sigma}_{\rm d}(r>r_{\rm c})={\bar \Sigma}_{\rm d}(r=r_{\rm c})$ \cite{hayashi05}.

The equations are numerically solved iteratively until the self consistency is attained
as $\delta Q < 2 \times 10^{-3}$, where
$\delta Q=\max_i\bigl\{ \bigl|Q_{\rm{new}}(r_i) - Q_{\rm{old}}(r_i)\bigr|/\bigl|Q_{\rm{new}}(r_i)\bigr|  \bigr\}$
and $Q$ stands for the pair potential and the self energies.
The larger the scattering rate $\Gamma_{\rm n}$ is, the slower the convergence becomes, especially in the unitary limit.
Therefore, we use an acceleration method \cite{Eschrig} for updating the pair potential and the self energies 
at each iterative step.



%% References


%\bibliographystyle{elsarticle-num}

\begin{thebibliography}{00}
\bibitem{shift}
The parameters $\sigma$ and $1/2\tau$ appearing in Refs.~\cite{preosti94,choi89,preosti96}
correspond to $\sigma=\sin^2\delta_0$ and $1/2\tau=\Gamma_{\rm n}$, respectively.
Here, $\delta_0$ and $\Gamma_{\rm n}$ are defined in the text.
\bibitem{fitting}
The zero-energy local density of states
can be fitted by the Gaussian function
$\exp(-r^2/c^2)$ 
in the vicinity of the vortex center
(see Refs.~\cite{kato01,nagai06}).
The half width at half maximum is expressed as
$\sqrt{\ln2} c = \sqrt{(\ln2)\xi_0 \xi_1(T)}$.
Here, $\xi_0$ and $\xi_1(T)$ are defined in the text.
\bibitem{log}
A deviation from the logarithmic temperature dependence
$1/\ln(T_{\rm c}/T) \sim 1/\ln[\xi_0/\xi_1(T)]$
of the zero-energy density of states
(see Refs.~\cite{KP,doettinger97,kato01}) corresponds to
that from the KP vortex core shrinkage.
It could be detected in the specific heat under a magnetic field
at low temperatures.
\bibitem{initial}
In this study, we used these initial values.
However, the stable (physical) solution of the Riccati equation can also be obtained
by using arbitrary initial values,
according to Ref.~\cite{nagai}.
\bibitem{self}
Numerical results indicate that $\Sigma_{\rm d}$ is pure imaginary.
We consider the radial line in the direction $\phi=0$, along which $\Sigma_{12}$ and $\Sigma_{21}$ are pure real.
Note $\Sigma_{\rm d}(-\omega_n)=[\Sigma_{\rm d}(\omega_n)]^*$ and $\Sigma_{12}(-\omega_n)=-[\Sigma_{21}(\omega_n)]^*$
because of Eq.~(\ref{eq:imp-self2}) and the general relations $g(-i\omega_n)=-[g(\omega_n)]^*$ and $f(-\omega_n)=[f^\dagger(\omega_n)]^*$.
Therefore, $\mathrm{Im}\Sigma_{\rm d}(-\omega_n)=-\mathrm{Im}\Sigma_{\rm d}(\omega_n)$
and $\mathrm{Re}\Sigma_{12}(-\omega_n)=-\mathrm{Re}\Sigma_{21}(\omega_n)$.
On the radial line for $\phi=0$, numerical results indicate $\mathrm{Re}\Sigma_{12}(-\omega_n)=\mathrm{Re}\Sigma_{12}(\omega_n)$, and therefore $\mathrm{Re}\Sigma_{21}(\omega_n)=-\mathrm{Re}\Sigma_{12}(\omega_n)$.
\bibitem{sign}
The sign difference in $\Sigma_{12}$ between here and
%Ref.~[8]
Ref.~\cite{sigrist91}
is inferred to be due to a coefficient difference in
the definition of the anomalous Green's function, which is physically irrelevant.
\end{thebibliography}



\end{document}